\documentclass[doublecol]{epl2}

\usepackage{graphicx}
\usepackage{epstopdf}
\usepackage{amssymb}
\usepackage{amsmath}
\usepackage{xspace}
\usepackage{bbold}

\newcommand{\ket}[1]{\ensuremath{|#1\rangle}\xspace}
\newcommand{\bra}[1]{\ensuremath{\langle #1|}\xspace}
\newcommand{\ps}[2]{\ensuremath{\langle #1|#2\rangle}\xspace}

\newcommand{\be}{\begin{equation}}
\newcommand{\ee}{\end{equation}}

\title{Euclidean matrix theory of random lasing in a cloud of cold atoms}

\author{A. Goetschy\inst{1} \and S.E. Skipetrov\inst{1}}
\shortauthor{A. Goetschy and S.E. Skipetrov}

\institute{
  \inst{1} Universit\'{e} Grenoble 1/CNRS, LPMMC UMR 5493, B.P. 166, 38042 Grenoble, France
}
\pacs{42.55.Zz}{Random lasers}
\pacs{42.25.Dd}{Wave propagation in random media}

\abstract{
We develop an \textit{ab initio} analytic theory of random lasing in an ensemble of atoms that both scatter and amplify light. The theory applies all the way from low to high density of atoms. The properties of the random laser are controlled by an Euclidean matrix with elements equal to the Green's function of the Helmholtz equation between pairs of atoms in the system. Lasing threshold and the intensity of laser emission are calculated in the semiclassical approximation. The results are compared to the outcome of the diffusion theory of random lasing.}

\begin{document}

\maketitle

\section{Introduction}

We call `random' a laser in which the feedback is provided by the multiple scattering of light on the random heterogeneities of the active medium and not by a well-defined external cavity \cite{cao05}. Recent theoretical models of random lasers rely on expansions of the laser field in terms of overlapping modes of `random cavities' formed by these heterogeneities \cite{deych05,tureci09}. Alternative approaches consist in solving Maxwell-Bloch equations numerically \cite{jiang00, conti08,andreasen10} or within the diffusion approximation \cite{letokhov68,wiersma98,perez09}. The latter has the advantage of yielding a simple criterion for the lasing threshold but it lacks rigorous justification and breaks down in the strong scattering regime \cite{cao05}.

In the present Letter we develop a new approach to the problem of random lasing that does rely neither on the expansion of the laser field in terms of cavity modes, nor on the diffusion approximation.
It is based on our recent analytic results for the `random Green's matrix' that describes propagation of light between scattering centers in a random medium \cite{goetschy11}.
To demonstrate the power of this new approach, we consider random lasing in an ensemble of a large number $N$ of identical atoms in free space, a problem of recent interest \cite{perez09,guerin10,savels05}. Dynamic equations for $N$ atoms that both scatter and amplify light are presented and analyzed in the semiclassical limit. We obtain analytic results for the lasing threshold and the average emitted intensity, thus achieving an important progress with respect to previous works on similar systems by Savels \textit{et al.} (who treated lasing in ensembles of $N \leq 5$ three-level atoms) \cite{savels05} and Froufe-P\'{e}rez \textit{et al.} (who dealt with $N \gg 1$ two-level atoms but in the diffusion approximation) \cite{perez09}. Our approach can be extended to deal with more `standard' random lasers in which scattering centers (`particles') are embedded in an amplifying homogeneous matrix \cite{goetschy12}.

\section{The model}

Consider a gas of $N$ three-level atoms at random positions $\mathbf{r}_i$ ($i = 1, \ldots ,N$) in free three-dimensional space. The atoms are subject to a strong external pump field of amplitude $\mathbf{E}_p(\mathbf{r})$, resonant with the transition from the ground state $\ket{a_i}$ to the upper auxiliary level $\ket{c_i}$ of each atom. They then rapidly decay to the upper level $\ket{b_i}$ of the laser transition at a rate $\Gamma_{cb} \gg \Gamma_{ba} = \Gamma \gg \Gamma_{ca}$. Interaction of atoms with the electromagnetic field which is near-resonant with the transition from $\ket{b_i}$ to $\ket{a_i}$ (energy difference $\hbar \omega_0$) can be described using standard approaches \cite{cohen92}, although certain subtleties should be treated with care. Relegating the derivation of dynamic equations for population imbalances $\hat{\Pi}_i=\ket{b_i}\bra{b_i}-\ket{a_i}\bra{a_i}$ and atomic lowering operators $\hat{S}^-_i = \ket{a_i}\bra{b_i}$ to a future publication \cite{goetschy12}, we present here their semiclassical approximation:
\begin{subequations}
\label{Langevin all}
\begin{eqnarray}
\frac{d{\Pi}_i}{dt} &=& -(1+W_i) \Pi_i+W_i-1
+ i\! \left[ S_i^* \Omega_i - \mathrm{c.c.} \right],\;\;\;\;\;
\label{Langevin Pi}
\\
\frac{d{S}_i}{dt} &=& \left[-\frac{i\omega_0}{\Gamma}-
\frac{1}{2}(1+W_i)\right] S_i
- \frac{i}{2} \Pi_i \Omega_i,
\label{Langevin S}
\end{eqnarray}
\end{subequations}
where time $t$ is in units of $\Gamma^{-1}$.
$\Pi_i=\bra{0} \hat{\Pi}_i \ket{0}$ and $S_i=\bra{0} \hat{S}^-_i \ket{0}$, with $\ket{0}$ the vacuum field state \cite{cohen92}, measure the difference between populations of levels $\ket{b_i}$ and $\ket{a_i}$ and the (induced) dipole moment of the atom $i$, respectively.
The pumping rate is given by $W_i = [\mathbf{d}_i \cdot \mathbf{E}_p(\mathbf{r}_i)]^2/
\hbar^2 \Gamma_{cb} \Gamma$ with $\mathbf{d}_i$ the dipole moment for the $\ket{c_i} \to \ket{a_i}$ transition. $\Omega_i$ is the $i$-th element of the vector $\mathbf{\Omega}={G}(\omega_0) \mathbf{S}$, with  $\mathbf{S}=(S_1, \ldots, S_N)$. It is proportional to the amplitude of the electric field at atom $i$ and is due to radiation of all other atoms.
The $N \times N$ non-Hermitian Green's matrix $G$ with elements
\be \label{Green matrix}
G_{ij}(\omega_0)=
(1-\delta_{ij})\frac{e^{ik_{0}|\mathbf{r}_i-
\mathbf{r}_{j}|}}{k_{0}|\mathbf{r}_i-\mathbf{r}_j|}
\ee
couples different atoms
($k_0 = \omega_0/c = 2\pi/\lambda_0$).
This matrix describes propagation of light between atoms and
belongs to the family of Euclidean random matrices defined by positions of $N$ points in the Euclidean space \cite{mezard99}.

Equations (\ref{Langevin all}) are derived in the scalar approximation for the electromagnetic field and assuming $\Gamma \ll \omega_0$, $c/R$ (with $R$ the size of the atomic cloud). They can be regarded as a generalization of the optical Bloch equation \cite{cohen92} to an ensemble of identical, incoherently pumped atoms. In the absence of coupling between atoms, they describe an isolated atom and have the stationary solution $\Pi_i = (W_i-1)/(W_i+1)$, $S_i = 0$, that shows that population inversion $\Pi_i > 0$ can be achieved for $W_i > 1$. This threshold for achieving population inversion, as well as the power broadening of the transition (the natural linewidth $\Gamma$ is increased by a factor $1 + W_i$), are due to sharing of the same ground state by the pump and the lasing transition. The dimensionless polarizability $\tilde{\alpha} = (k_0^3/4\pi) \alpha = S_i(\omega)/\Omega_i(\omega)$ of an atom pumped at a rate $W$ follows from eqs.\ (\ref{Langevin all}) (see also ref.\ \cite{savels05}):
\be
\tilde{\alpha}(\omega, W) = \frac{W-1}{W+1}\;\frac{\Gamma/2}{(\omega-\omega_0)
+i(1+W)\Gamma/2}.
\label{polarizability}
\ee

\section{Lasing threshold}

\begin{figure*}
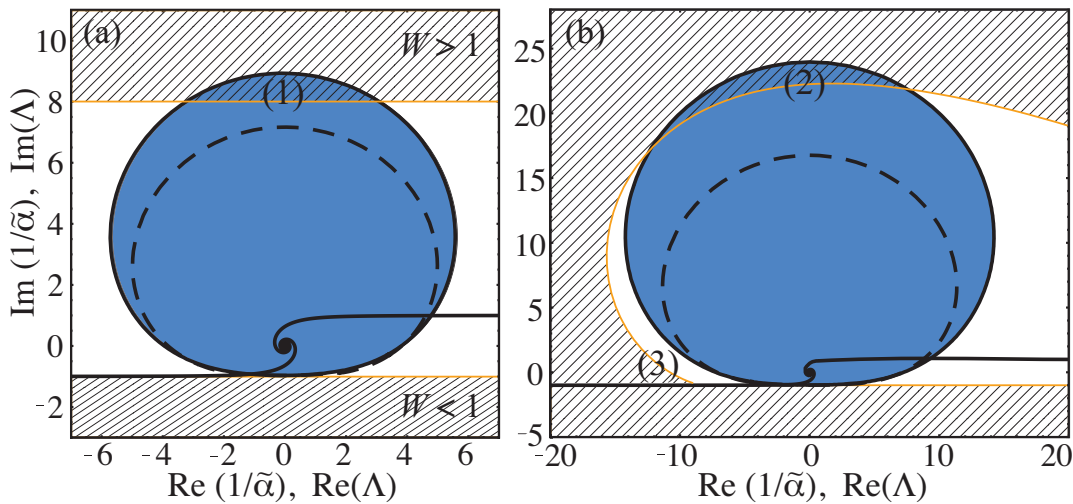

\centering{
\onefigure[angle=0,width=0.8\textwidth]{contour_threshold}
\caption{
The domain $\mathcal{D}_\alpha$ (hatched) spanned by
$1/\tilde{\alpha}$ and the domain $\mathcal{D}_\Lambda$ (blue area delimited by the solid line) occupied by the eigenvalues $\Lambda$ of the random Green's matrix (\ref{Green matrix}). (a) Incoherent gain $\tilde{\alpha}(\omega, W)$, see eq. (\ref{polarizability}). (b) Coherent Mollow gain $\tilde{\alpha}(\delta, \Delta, \Omega)$ with $\Delta=\Gamma$ \cite{mollow72,boyd,perez09}.  Lasing occurs when $\mathcal{D}_\alpha$ and $\mathcal{D}_\Lambda$ overlap: regions (1), (2), (3). The borderline of $\mathcal{D}_\Lambda$ is given by eq.\ (\ref{threshold3}) with the optical thickness $b_0=40$ in (a) and $b_0=140$ in (b). The dashed lines show the borderline of $\mathcal{D}_\Lambda$ following from the diffusion approximation [eq.\ (\ref{diffusion})].
\label{fig1}
}}
\end{figure*}

When coupling between different atoms is at work, the stationary solution $S_i = 0$ of eqs.\ (\ref{Langevin all}) may lose its stability for a sufficiently strong pump. Following standard semiclassical laser theories \cite{haken85}, we will associate this instability with reaching the lasing threshold. The stability analysis of nonlinear eqs.\ (\ref{Langevin all}) shows that for uniform pump $W_i = W$ lasing starts when at least one of the eigenvalues $\Lambda_k$ of ${G}$ satisfies
\be
\frac{2W}{1+W} \mathrm{Im}\Lambda_k > (1+W)+\mathrm{Im} \Lambda_k.
\label{threshold}
\ee
The left-hand side of this condition can be regarded as gain that depends on both the pumping rate $W$ and scattering (through $\Lambda_k$), whereas the right-hand side contains pump-dependent losses due to spontaneous emission ($1+W$) and leakage out of the system ($\mathrm{Im} \Lambda_k$). As counterintuitive as it may seem, it follows from eq.\ (\ref{threshold}) that random lasing takes place when $\mathrm{Im} \Lambda_k$ (that quantifies losses due to open boundaries in the absence of pump) \textit{exceeds} $(1+W)^2/(W-1)$ and $W > 1$.

We now propose a less technical and physically transparent derivation of the threshold condition (\ref{threshold}) which, in addition, can be generalized to an arbitrary pumping scheme. On the one hand, the electric field at atomic positions is $\mathbf{\Omega}(\omega)={G}(\omega) \mathbf{S}(\omega)$, whereas on the other hand, the linear response of atoms to the field implies $\mathbf{S}(\omega)={\cal A}(\omega)
\mathbf{\Omega}(\omega)$, where ${\cal A}$ is the diagonal matrix ${\cal A} = \textrm{diag}[\tilde{\alpha}_i]$ and $\tilde{\alpha}_i$ is the dimensionless polarizability of atom $i$. It then follows immediately that the linear description breaks down and lasing starts when one of the eigenvalues of the matrix product ${G}(\omega){\cal A}(\omega)$ is equal to one. For atoms in free space, $G(\omega)$ can be safely replaced by $G(\omega_0)$. Therefore, for identical atoms and uniform pumping ($\tilde{\alpha}_i = \tilde{\alpha}$) the threshold condition reduces to
\be
\Lambda_k = \frac{1}{\tilde{\alpha}}.
\label{threshold2}
\ee
This equation illustrates that laser threshold results from an interplay of single-atom properties (described by the polarizability $\tilde{\alpha}$) and geometry-dependent collective effects (quantified by the random eigenvalues $\Lambda_k$ of the Green's matrix $G$). If we substitute the polarizability (\ref{polarizability}) into eq.\ (\ref{threshold2}), we recover eq.\ (\ref{threshold}). But eq.\ (\ref{threshold2}) is more general than eq.\ (\ref{threshold}) and is not restricted to lasing in a system of three-level atoms. We can also apply it, for example, to an ensemble of two-level atoms (resonant frequency $\omega_0$) in the field of a strong near-resonant coherent pump (frequency $\omega_0 + \Delta$, Rabi frequency $\Omega$). When illuminated by a weak probe light at a frequency $\omega_0 + \Delta + \delta$, each atom behaves as if it had an effective polarizability $\tilde{\alpha}(\delta, \Delta, \Omega)$ with a long but manageable explicit expression that can be found in Refs.\ \cite{mollow72,boyd,perez09}. Optical gain in such a system is sometimes referred to as `Mollow gain' \cite{mollow72}.

An easy way to visualize the threshold condition (\ref{threshold2}) is to draw the two-dimensional domain $\mathcal{D}_\Lambda$ occupied by the eigenvalues of $G$ and the region $\mathcal{D}_\alpha$ spanned by $1/\tilde{\alpha}$ when its free parameters --- $\omega$ and $W$ in the case of eq.\ (\ref{polarizability}) --- are varied, on the complex plane. Random lasing takes place when $\mathcal{D}_\Lambda$  and $\mathcal{D}_\alpha$ touch (threshold) or overlap.
This is illustrated in fig.\ \ref{fig1} for $N \gg 1$ atoms in a sphere of radius $R \gg \lambda_0$. In this figure, we adjusted the parameters for the random laser to be slightly above threshold: $\mathcal{D}_\Lambda$ and $\mathcal{D}_\alpha$ barely overlap. Whereas $\mathcal{D}_\alpha$ is easy to determine when $\tilde {\alpha}$ is known as a function of its parameters, finding $\mathcal{D}_\Lambda$ is much less trivial. Here we make use of our recent results for the eigenvalue distribution of the Green's matrix (\ref{Green matrix}) in the limit of large $N$ \cite{goetschy11}. The distribution and the boundary of its support $\mathcal{D}_\Lambda$ on the complex plane depend on two dimensionless parameters: the number of atoms per wavelength cubed $\rho\lambda_0^3$  and the on-resonance optical thickness $b_0 = 2R/\ell$, where $\rho$ is the number density of atoms and $\ell=k_0^2/4\pi\rho$ is the on-resonance scattering mean free path in the absence of the pump. The eigenvalue domain $\mathcal{D}_\Lambda$ consists of two parts: a (roughly circular) `bulk' and a pair of spiral branches \cite{goetschy11,skipetrov11}. Depending on the particular model of atomic polarizability $\tilde{\alpha}$, either the bulk or the branches may touch ${\cal D}_{\alpha}$, as we now discuss.

\begin{figure*}[t]
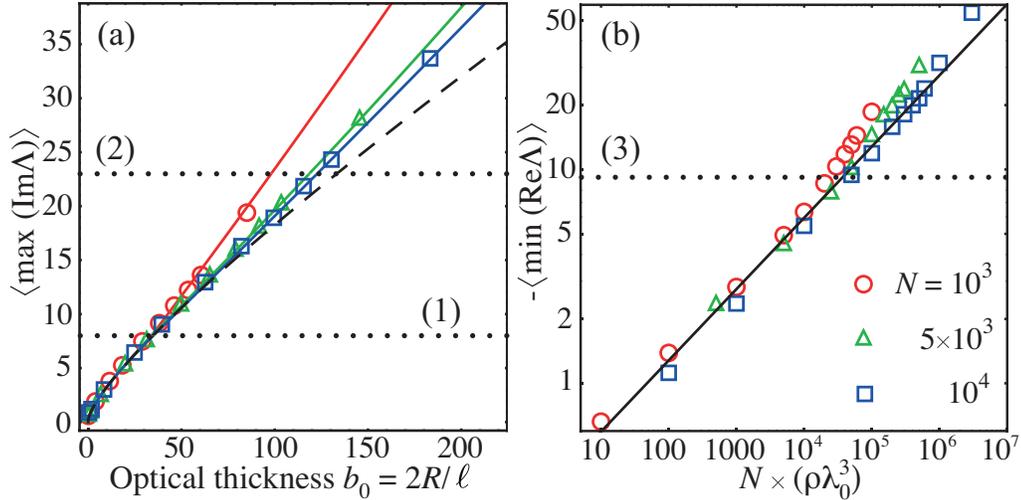

\centering{
\onefigure[angle=0,width=0.75\textwidth]{maxIm_minRe}
\caption{
Mean maximum value of the imaginary part (a) and mean minimum value of the real part (b) of eigenvalues $\Lambda$ of the $N \times N$ random Green's matrix (\ref{Green matrix}). Our analytic results (solid lines) are compared with numerical simulations for three different matrix sizes $N$ (symbols). The horizontal dotted lines indicate the absolute lasing thresholds: (1) for the incoherent gain [eq.\ (\ref{polarizability})] and (2) and (3) for the coherent Mollow gain \cite{perez09,mollow72,boyd}. In (a), analytic results depend on both $b_0$ and $\rho\lambda_0^3$ \cite{goetschy11}, except for $\rho\lambda_0^3\lesssim10$ when they reduce to eq.\ (\ref{threshold3}) with $\tilde{\alpha}$ replaced by $1/i\left<\textrm{max}(\textrm{Im}\Lambda)\right>$ (dashed line). The analytic result $\left<\textrm{min}(\textrm{Re}\Lambda)\right>=
-\Gamma(2/3)(N\rho\lambda_0^3/12\pi^2)^{1/3}$ [solid line in (b)] is valid for $\rho\lambda_0^3\lesssim10$.
\label{fig2}
}}
\end{figure*}

We first focus on the lasing threshold due to the bulk of eigenvalues. Combining the analytic equation for the borderline of $\mathcal{D}_{\Lambda}$ at low density $\rho\lambda_0^3 \lesssim 10$ \cite{goetschy11} and eq.\ (\ref{threshold2}) results in a threshold condition that depends on the optical thickness $b_0$ but not on the density $\rho\lambda_0^3$:
\be
\frac38 b_0 |\tilde{\alpha}|^2 h\left( \frac12 b_0\textrm{Im}\tilde{\alpha} \right) =1,
\label{threshold3}
\ee
where  $h(x)=[3-6x^2+8x^3-3(1+2x)e^{-2x}]/6x^4$. Note that for both gain mechanisms considered in this Letter, the threshold condition (\ref{threshold3}) involves the eigenvalue with the largest imaginary part, as can be seen from fig.\ \ref{fig1}. We calculated $\left<\textrm{max}(\textrm{Im}\Lambda)\right>$ based on our non-Hermitian random matrix theory \cite{goetschy11} and found excellent agreement with numerical results, see fig.\ \ref{fig2}(a). It is quite remarkable that the agreement is present at all values of parameters, including high densities $\rho \lambda_0^3 \gg 1$ that were necessary to reach large optical thicknesses $b_0 \gg 1$ in numerical calculations with moderate $N \leq 10^4$. Because it is $\left<\textrm{max}(\textrm{Im}\Lambda)\right>$ that controls the laser threshold, we conclude that our theory applies to random lasing all the way from weak ($\rho\lambda_0^3 \ll 1$) to strong ($\rho\lambda_0^3 \gg 1$) scattering regime.

\begin{figure}
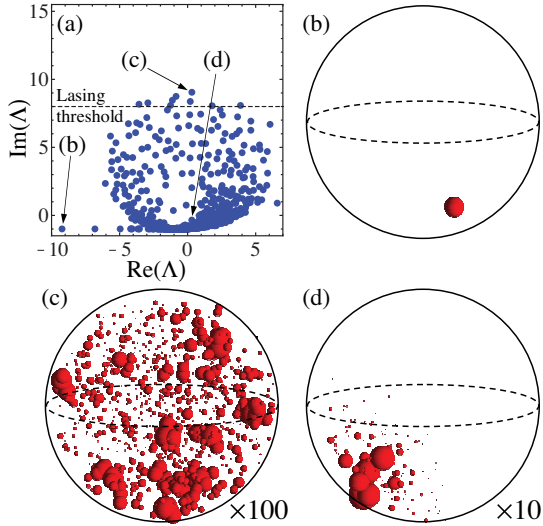

\centering{
\onefigure[width=0.42\textwidth]{modes}
\caption{
(a) Eigenvalues $\Lambda$ of a single random realization of the Green's matrix $G$ (dots) for a cloud of optical thickness $b_0 = 40$, composed of $N = 10^3$ atoms. (b)--(d) Intensities $|R_k^i|^2$ corresponding to the mode in the subradiant branch, localized on a pair of atoms (b), the mode with the largest $\mathrm{Im} \Lambda$ (c) and the mode corresponding to the smallest $|\Lambda|$ (d). A mode $\mathbf{R}_k = \{ R_k^1, R_k^2, \ldots, R_k^N \}$ is represented by spheres centered at positions of atoms
$\mathbf{r}_i$ and having radii equal to $1 \times$ (b), $100 \times$ (c), and $10 \times |R_k^i|^2$ (d).}
\label{fig25}
}
\end{figure}

It is interesting to compare the threshold condition (\ref{threshold3}) with the one obtained in the diffusion approximation. The latter amounts to solve the diffusion equation for the average intensity of light with gain included as a negative absorption \cite{letokhov68,wiersma98,perez09}. The threshold is reached when the solution diverges. This yields the following threshold condition \cite{perez09,goetschy11}\footnote{We use the value of the extrapolation length $z_0 = 2/3$ instead of $z_0 = 0.71$ in Ref.\ \cite{perez09}.}:
\be
\frac{\sqrt{3}}{2\pi} b_0 |\tilde{\alpha}|
\sqrt{|\tilde{\alpha}|^2 - \mathrm{Im} \tilde{\alpha}}
\left[1 + \frac{1}{1 + \frac34 b_0 |\tilde{\alpha}|^2} \right] = 1.
\label{diffusion}
\ee
This equation is similar to our result (\ref{threshold3}) at large optical thickness $b = b_0 |\tilde{\alpha}|^2 \gg 1$ but deviates significantly at $b \lesssim 1$, as can be seen from fig.\ \ref{fig1}. Consequently, the predictions of eq.\ (\ref{diffusion}) for the laser threshold (that is reached at $b < 1$, see fig.\ \ref{fig1}) turn out to be inaccurate.
In particular, our eq.\ (\ref{threshold3}) predicts that the minimum on-resonance optical thicknesses required for random lasing are $b_{0\mathrm{cr}} \simeq 35$ for the incoherent and $b_{0\mathrm{cr}} \simeq 110$ for the coherent pump. This is significantly less than 50 and 200, respectively, following from eq.\ (\ref{diffusion}).
Analysis of the right eigenvectors $\mathbf{R}_k$ (modes) of the matrix $G$ shows that at all densities $\rho\lambda_0^3$, the mode that reaches the threshold first is extended over the whole atomic cloud [see fig.\ \ref{fig25}(c)], even when the system may support localized modes as well [see modes (b) and (d) in fig.\ \ref{fig25}]. This is specific for the models considered here in which, in particular, scattering and gain are due to the same atoms, and in contrast with systems where gain and scattering are independent and (pre-)localized modes may be better candidates for lasing \cite{cao00,apalkov04} (see ref.\ \cite{andreasen10} for a recent review).

In the high-density limit $\rho\lambda_0^3\to \infty$, the eigenvalues of $G$ that have large imaginary parts collapse on a line described by a simple equation \cite{goetschy11} which, combined with eq.\ (\ref{threshold2}), yields the lasing threshold condition for a continuous medium with a refractive index
$n(\tilde{\alpha})=(1+\tilde{\alpha}\rho\lambda_0^3/2\pi^2)^{1/2}$:
\be
\left|\frac{n(\tilde{\alpha})-1}{n(\tilde{\alpha})
+1}\right|^2\left|e^{4in(\tilde{\alpha})k_0R}\right|=1.
\label{threshold4}
\ee
In this limit the problem looses its statistical nature and the random laser turns into a `standard' laser with the feedback due to (partial) reflections at the boundaries of a homogeneous amplifying medium.


\begin{figure*}[t]
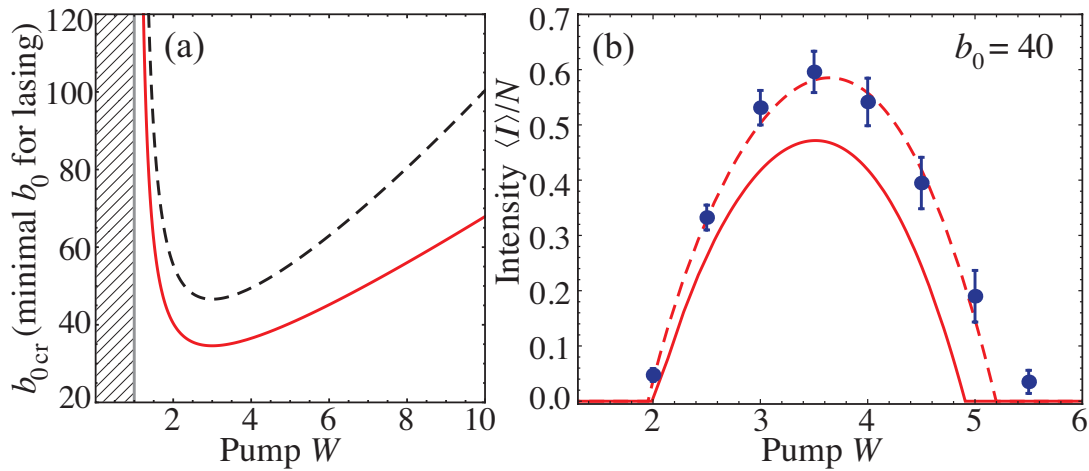

\centering{
\onefigure[angle=0,width=0.8\textwidth]{b0_intensity_vsW}
\caption{
(a) The minimal (critical) optical thickness $b_{0\mathrm{cr}}$ necessary for lasing, following from our Euclidean matrix theory (\ref{threshold3}) (solid line) and from the diffusion approximation (\ref{diffusion}) (dashed line). (b) The average stationary intensity $\langle I \rangle$ at $b_0 = 40$ obtained from the numerical solution of eqs.\ (\ref{Langevin all}) for $N = 10^3$ after averaging over 10 random configurations of atoms (symbols). For each configuration and each $W$, we averaged the numerical solution $I(t)$ over $t = (200$--$250)\Gamma^{-1}$. The analytic solution (\ref{intensity}) is shown by the solid line. The dashed line corresponds to $y_0 = 9.15$ instead of 8.93 for the solid line.
\label{fig3}
}}
\end{figure*}

Let us now analyze the role of the spiral branches of $\mathcal{D}_\Lambda$.
As we illustrate in fig.\ \ref{fig25}(b), the eigenvalues belonging to these spirals correspond to eigenvectors (modes) localized on pairs of very close points $|\mathbf{r}_i - \mathbf{r}_j| \ll \lambda_0$ \cite{goetschy11,skipetrov11}.
These are the super- and sub-radiant states of a pair of atoms. For the uniform incoherent gain (\ref{polarizability}), the branches do not overlap with $\mathcal{D}_\alpha$ [fig.\ \ref{fig1}(a)], whereas the lower, `subradiant' branch overlaps with $\mathcal{D}_\alpha$ for the coherent Mollow gain [fig.\ \ref{fig1}(b), region (3)]. Thus, in the latter case the solution $S_i = 0$ of eqs.\ (\ref{Langevin all}) may lose its stability due to the eigenvalue with the smallest real part belonging to this branch. We calculated $\left<\textrm{min}(\textrm{Re}\Lambda)\right>$ from the probability to find two points separated by a distance $\Delta r > x R$: $p = [1-x^3(1-9x/16+x^3/32)]^{N(N-1)/2}$. The result scales with $(N\rho\lambda_0^3)^{1/3}$, in good agreement with numerical simulations as long as $\rho\lambda_0^3\lesssim10$ [fig. \ref{fig2}(b)].
When $(-\mathrm{Re} \Lambda)$ exceeds a critical value, a pair of closely located atoms on which the eigenvector (mode) associated with the eigenvalue $\Lambda$ is localized, starts to emit coherent light. On average, the threshold for this effect is given by the condition
\be
\label{branch}
-\Gamma(2/3) \left( \frac{N\rho\lambda_0^3}{12\pi^2} \right)^{1/3}
= \frac{1}{\tilde{\alpha}} + i,
\ee
where $\Gamma(x)$ is the Gamma function.
Formally, this emission of light by a pair of pumped atoms may be called `laser', especially given the fact that one-atom cavity lasers \cite{mckeever03} and few-atom random lasers \cite{savels07} were already discussed in the literature. It is very different from the collective laser mechanism leading to eqs.\ (\ref{threshold3}) and (\ref{threshold4}) and associated with eigenvectors extended over the whole atomic system. Whereas eqs.\ (\ref{threshold3}) and (\ref{threshold4}) are good estimates of the threshold even for a \textit{single\/} atomic configuration, the threshold for light emission by a pair of atoms is expected to fluctuate strongly around its typical value given by eq.\ (\ref{branch}). In addition, this emission is strongly affected by quantum effects not included in our analysis. In fact, quantum fluctuations wash out the sharp threshold obtained in the semiclassical framework \cite{goetschy12}.


\section{Laser intensity}

Let us now study the dynamics of laser emission slightly above threshold. In the vicinity of threshold, $\Pi_i$ can be adiabatically eliminated from eqs.\ (\ref{Langevin all}) \cite{haken85}. Keeping only the lowest-order nonlinear terms, we obtain an equation for the dimensionless field amplitude $\mathbf{\Omega}=G \mathbf{S}$:
\be
\label{adiabatic field}
\frac{d\mathbf{\Omega}}{dt} = -\left[i \left( \frac{\omega_0}{\Gamma} + GA - GC|\mathbf{\Omega}|^2 \right)+ G BG^{-1} \right] \mathbf{\Omega},
\ee
where we introduced $N\times N$ diagonal matrices $A = \textrm{diag}[(W_i-1)/(W_i+1)]/2$, $B=\textrm{diag}[W_i+1]/2$,  $C=\textrm{diag}[(W_i-1)/(W_i+1)^3]$, and $|\mathbf{\Omega}|^2=\textrm{diag}[|\Omega_i|^2]$.
Note that the eigenvectors of the linear kernel of eq.\ (\ref{adiabatic field}) coincide with the right eigenvectors $\mathbf{R}_k$ of the matrix $G$, that play the role of eigenmodes of the `cold cavity', only if the pump is uniform: $W_i = W$. Restricting further consideration to the latter case, we introduce an ansatz $\mathbf{\Omega}=\sum_k a_k(t)e^{-i\omega_k t} \mathbf{R}_k$ \cite{haken85} and obtain rate equations for mode intensities $I_k = |a_k|^2 \ps{\mathbf{R}_k}{\mathbf{R}_k}$:
\be
\frac{dI_k}{dt}=-2 \kappa_k I_k + \sum_n \mathcal{W}_{nk} I_n I_k,
\label{rate equations}
\ee
where
$2\kappa_k = \Gamma [(W+1)-\textrm{Im}\Lambda_k (W-1)/(W+1)]$, $\mathcal{W}_{nk} = -4 \Gamma \textrm{Im}(\Lambda_k\gamma_{nk})(W-1)/(W+1)^3$, and a free-running situation (no phase locking) is assumed.  The threshold for the mode $k$ is given by the condition $\kappa_k = 0$ and depends only on the eigenvalue $\Lambda_k$, whereas the mode competition above the threshold involves the overlap of eigenvectors $\gamma_{nk}=\sum_{i=1}^N(R_k^i)^2|R_n^i|^2/\sum_{i=1}^N|R_n^i|^2$. It is worth noting that although rate equations similar to eq.\ (\ref{rate equations}) appeared in previous works on random lasers \cite{deych05}, loss rates $\kappa_k$ and nonlinear couplings $\mathcal{W}_{nk}$ were most often assumed to follow from \textit{ad hoc} random matrix models, except in one-dimensional systems where they could be calculated with a reasonable effort. We, in contrast, provide explicit general expressions for these quantities and show that they are determined by the eigenvalues $\Lambda_k$ and eigenvectors $\mathbf{R}_k$ of the random Green's matrix. The link between $\kappa_k$, $\mathcal{W}_{nk}$ and $\Lambda_k$, $\mathbf{R}_k$ is independent of the geometry or dimensionality of the problem.

At low atomic density $\rho\lambda_0^3 \lesssim 10$, $R_k^i$ behave almost as independent Gaussian random variables and thus $N \langle \gamma_{nk} \rangle \simeq 1 + 2\delta_{nk}$. This allows us to calculate the average intensity in the stationary regime $\langle I \rangle = \langle \sum_{i=1}^N |\Omega_i|^2 \rangle \simeq \langle
\sum_k I_k \rangle$:
\be
\langle I \rangle = N \frac{(1+W)^2}{4} \left[
1 - \frac{(W+1)^2}{W-1} \frac{1}{y_0} \right],
\label{intensity}
\ee
where $y_0$ is the solution of $y_0^2 = (3/8) b_0 h(-b_0/2y_0)$. This result is in good agreement with the numerical solution of eqs.\ (\ref{Langevin all}), as we show in  fig.\ \ref{fig3}(b).
The average number of lasing modes, i.e. of eigenvectors $\mathbf{R}_k$ of $G$ that have non-vanishing amplitudes $a_k(t)$ in the expansion of $\mathbf{\Omega}$, in the long-time limit, is $\left<N_L\right> = (2N)^{1/2} [(W-1)/(W+1)^2-1/y_0]^{1/2}y_0/(y_0+1)$. Note that in the absence of mode competition, $\langle I \rangle$ and $\langle N_L \rangle$ would scale as $N^2$ and $N$, respectively.

An interesting feature of lasing in a cloud of cold atoms illustrated by fig.\ \ref{fig3}(b) is the halt of lasing at too strong pumps. This can be easily understood by noting that random lasing requires both amplification and scattering to be sufficiently strong, and that both of these important ingredients are provided by the same atoms. At low pump [$W \lesssim 2$ in fig.\ \ref{fig3}(b)], the scattering is strong, but the amplification is not enough to lase. In contrast, when the pump is strong ($W \gtrsim 5$), the scattering strength decreases because the atomic transition starts to be saturated, and lasing stops.

\section{Possible extensions}

The results presented above concern the simplest case of random lasing in an ensemble of atoms in free space, in the scalar approximation and under a spatially uniform pump. This allows us to demonstrate the power of the Euclidean matrix approach which, in this case, yields explicit analytical results for the lasing threshold [eqs.\ (\ref{threshold3}), (\ref{threshold4}) and (\ref{branch})] and the laser intensity [eq.\ (\ref{intensity})]. However, the approach is not limited to this simplest case and can be extended to take into account additional complications that may arise in an experiment. If, for example, the pump is nonuniform, finding the lasing threshold will require diagonalization of the matrix $GA - i GBG^{-1}$ instead of $G$, as follows from eq.\ (\ref{adiabatic field}). For a complex distribution of pump, this can be realized only numerically. Another important situation is lasing in an ensemble of passive scatterers embedded in an amplifying matrix with an amplification coefficient $\mu$. In this case, one has to put $W = 0$ and study the eigenvalues of the Green's matrix
$G_{ij}(\omega_0)=
(1-\delta_{ij}) \exp[(ik_{0}+ \mu)|\mathbf{r}_i-
\mathbf{r}_{j}|]/k_{0}|\mathbf{r}_i-\mathbf{r}_j|$, which can be done analytically \cite{goetschy12}. Finally, taking into account the vector nature of light requires using the dyadic Green's function instead of the scalar one and working with $3N \times 3N$ matrices. The threshold condition is still given by eq.\ (\ref{threshold2}) and at low atomic densities considered in this work, the analysis of the threshold due to the bulk of eigenvalues can be reduced to the scalar approximation. However, the spiral branches of the distribution (see fig.\ \ref{fig1}) are sensitive to the vector nature of light and thus the subradiant laser threshold (\ref{branch}) will be modified.


\section{Conclusion}

Quite remarkably, the general framework developed in this paper and, in particular, the threshold condition (\ref{threshold2}), apply for any dimensionality of space, any atomic polarizability, any number and concentration of atoms, and any form of the Green's matrix $G(\omega)$ that, in particular, can account for an external cavity and amplification or absorption of light in the space between the atoms. This opens a way to treat various random laser systems analytically without strong approximations or heavy numerics.

Application of the Euclidean random matrix theory to the problem of random lasing in a cloud of cold atoms allowed us to find the lasing threshold without relying on the diffusion approximation or transport theory, as well as to study the intensity of laser emission beyond threshold. We predict the possibility of random lasing in a cloud of cold atoms for on-resonance optical thickness exceeding 35 for three- and 110 for two-level atoms (`Mollow laser'). We show that mode competition plays an important role in the random laser and leads to the scaling of the number of modes participating in lasing with $\sqrt{N}$ (instead of $N$ in the absence of mode competition), where $N$ is the number of atoms. At the same time, the laser intensity scales with $N$ (instead of $N^2$).

Finally, our approach based on Euclidean random matrix theory may be also applied to understand the role of disorder in other collective effects that take place in ensembles of cold atoms, such as the collective spontaneous emission \cite{svid10} or the radiative force experienced by an optically thick atomic sample \cite{bien10}.

\acknowledgments
This work was supported by the French ANR (project no. 06-BLAN-0096 CAROL). We thank B. van Tiggelen and D. Basko for carefully reading the manuscript.

\end{document}